\documentclass[conference]{IEEEtran}
\IEEEoverridecommandlockouts

\usepackage{cite}
\usepackage{amsmath,amssymb,amsfonts}
\usepackage{algorithmic}
\usepackage{graphicx}   
\usepackage{textcomp}
\usepackage{xcolor}
\usepackage{array}
\usepackage{booktabs}
\usepackage{multirow}
\usepackage{float}
\usepackage{comment}
\usepackage{subfigure}
\usepackage{url}  

\graphicspath{{Images/}}

\title{FCT O-RAN: Design and Deployment of a Multi-Vendor End-to-End Private 5G Testbed}

\author{
    Amogh PC, Nagamuthu Vignesh, Pei Yiyang, Neelakantam Venkatarayalu,\\
    Pedro Henrique Amorim Rezende, Shyam Babu Mahato, and Sumei Sun\\
    Future Communications Translation Lab (FCTLab) \\
    Singapore Institute of Technology, Singapore \\
    \texttt{\{amogh.pc, nagamuthu.vignesh, yiyang.pei, n.venkat,}\\
    \texttt{pedrohenrique.amorimrezende, shyam.mahato, sumei.sun\}@singaporetech.edu.sg}
}

\begin{document}

\maketitle

\begin{abstract}

The transformation of 5G networks into software-defined, agile, intelligent and programmable architectures necessitates a paradigm shift in deployment strategies. To deliver superior performance and surpass traditional systems, public and private 5G networks must adopt software-centric cloud native frameworks that enable flexibility through tailored configurations and optimized deployment approaches. In Singapore, the Infocomm Media Development Authority (IMDA) and the National Research Foundation Singapore (NRF) launched the Future Communications Research and Development Programme (FCP) to advance the nation's communications and connectivity landscape. At the core of this initiative is the Future Communications Translation Lab (FCT) at the Singapore Institute of Technology (SIT), which focuses on advancing 5G technologies to higher readiness levels, facilitating their adoption across various industries. A key component is the deployment of FCT O-RAN, a state-of-the-art multi-vendor private 5G platform. The setup includes a 5G core network powered by Microsoft Affirmed and ENEA, O-RAN Centralized and Distributed Units from Radisys. Indoor Remote Units are deployed with Foxconn, while outdoor RUs are deployed with Benetel. To optimize the deployment of remote units, a digital twin was created using Wireless InSite, and performance evaluations were conducted for both the digital twin and the private 5G deployment. Smartphones equipped with QualiPoc were used to measure network performance. The testbed demonstrated effective performance with optimized bandwidth allocations for both indoor and outdoor environments. In the indoor setup, utilizing 50 MHz of bandwidth, a downlink throughput of 713 Mbps and an uplink throughput of 66 Mbps were achieved. Meanwhile, the outdoor setup, utilizing 40 MHz of bandwidth, achieved a downlink throughput of 371 Mbps and an uplink throughput of 55 Mbps.

\end{abstract}


\section*{Introduction}

With 5G advancing to the forefront of wireless communication, it signifies a fundamental shift toward open, intelligent, and software-driven architectures. This evolution is anchored on three core principles: \begin{itemize} \item \textbf{Disaggregation}: Breaking the RAN into modular components such as the Central Unit (CU), Distributed Unit (DU), and Radio Unit (RU) to enable flexibility and scalability. \item \textbf{Softwarization}: Transitioning from hardware-specific functionalities to adaptable, software solutions~\cite{amogh2024defending}. \item \textbf{Intelligent Control}: Leveraging AI-powered, closed-loop systems~\cite{villa2024open} to optimize and automate RAN operations for greater efficiency. \end{itemize}

Open Radio Access Network (O-RAN)~\cite{marinova2024intelligent} architectures epitomize these principles, fundamentally transforming the telecom ecosystem. By enabling modularity and programmability, O-RAN facilitates seamless integration of components from multiple vendors through open and standardized interfaces. This fosters innovation, competition, and flexibility, allowing networks to be tailored to diverse and evolving requirements. The benefits of this transformation are driven by a diversified telecom ecosystem comprising open-source initiatives, specialized vendors, and innovative collaborations. Efforts such as the AI-RAN Alliance, which focuses on native AI integration for advanced network optimization, exemplify the forward-looking approach to 6G RAN evolution.

Additionally, global initiatives such as the Next G Alliance, IMT-2030, the Telecom Infra Project (TIP), and the 5G Public-Private Partnership (5GPPP) promote open and collaborative strategies for the deployment and evolution of 5G networks, as well as the transition to 6G. At the forefront of Singapore’s transformation in communications and connectivity is FCT O-RAN, an end-to-end 5G platform developed as part of ongoing initiatives at the Future Communications Translation Lab (FCT) at the Singapore Institute of Technology (SIT). This platform establishes a high-bandwidth private 5G network that seamlessly operates across both indoor and outdoor environments.

The development of this testbed presents significant challenges in integrating multi-vendor systems. Rigorous testing, troubleshooting, and close coordination with key stakeholders and other industry partners. The FCT O-RAN integrates technologies from leading global vendors to create a robust and open infrastructure. This end-to-end 5G testbed is powered by a 5G core network utilizing Microsoft affirmed~\cite{affirmed_networks} and ENEA's~\cite{enea_website} cloud-native functions. Radisys~\cite{radisys_website} provides the O-RAN CU and DU, adhering to open standards to ensure seamless interoperability and flexibility. The physical layer is powered by Intel FlexRAN~\cite{intel_flexran}.

To ensure seamless coverage across diverse environments, the testbed includes both indoor and outdoor radio units (RUs), with indoor RUs procured from Foxconn~\cite{foxconn_website} and outdoor RUs procured from Benetel~\cite{benetel_website}. To optimize the placement of these RUs, we employed Wireless InSite~\cite{wireless_insite} for creating a digital twin to optimize RU placement, enhance coverage, and improve overall 5G network performance.

\color{black}

Our contributions in this paper are summarized as follows:

\begin{itemize}
    \item \textbf{Development of a Multi-Vendor Private 5G Testbed:}  
    We present the deployment of FCT O-RAN, integrating 5G components from Microsoft Affirmed, ENEA, Radisys, Foxconn, and Benetel. This private 5G network operates in the 3400--3450 MHz (FR1, n78) band.

    \item \textbf{Cloud-Native 5G O-RAN Architecture:}  
   We implement and adhere to a cloud-native 5G network architecture based on the principles of the European Telecommunications Standards Institute Management and Orchestration (ETSI MANO), integrating the O-RAN Radisys Trillium 5G NR gNB solution. For the 5G core, we utilize ENEA's Integrated 5G Data Management and Policy Control system. Additionally, we deploy Microsoft Affirmed Network Functions to enable other essential 5G core capabilities.

    \item \textbf{Optimization of RF Placement Using Wireless InSite Digital Twin:}  We utilize a digital twin created with Wireless InSite to optimize the placement of Remote Units, thereby enhancing coverage and performance. We validate the testbed through walk tests with UEs equipped with QualiPoc$^\text{TM}$~\cite{rohde_schwarz_qualipoc_android}, measuring throughput, Signal-to-Interference-plus-Noise Ratio (SINR), and Reference Signal Received Power (RSRP), and comparing the results with the predictions of the digital twin.

\end{itemize}

The rest of the paper is organized as follows: Section~\ref{sec:deployment_sit_infrastructure} presents an overview of the deployment detailing the infrastructure and configurations for indoor and outdoor. Section~\ref{sec:multi_vendor_fct_oran_architecture} describes the Multi-Vendor FCT O-RAN Cloud-Native Architecture, focusing on the integration of key components and cloud-native functionalities. Section~\ref{sec:experimental_results} discusses the experimental results, including the digital twin setup, real-world measurements, and their comparison. Finally, Section~\ref{sec:conclusion} concludes the paper by summarizing the contributions and future work.

\section{Deployment at SIT: Overview and Infrastructure}
\label{sec:deployment_sit_infrastructure}

\begin{figure}[h!]
    \centering
    \includegraphics[scale=0.18]{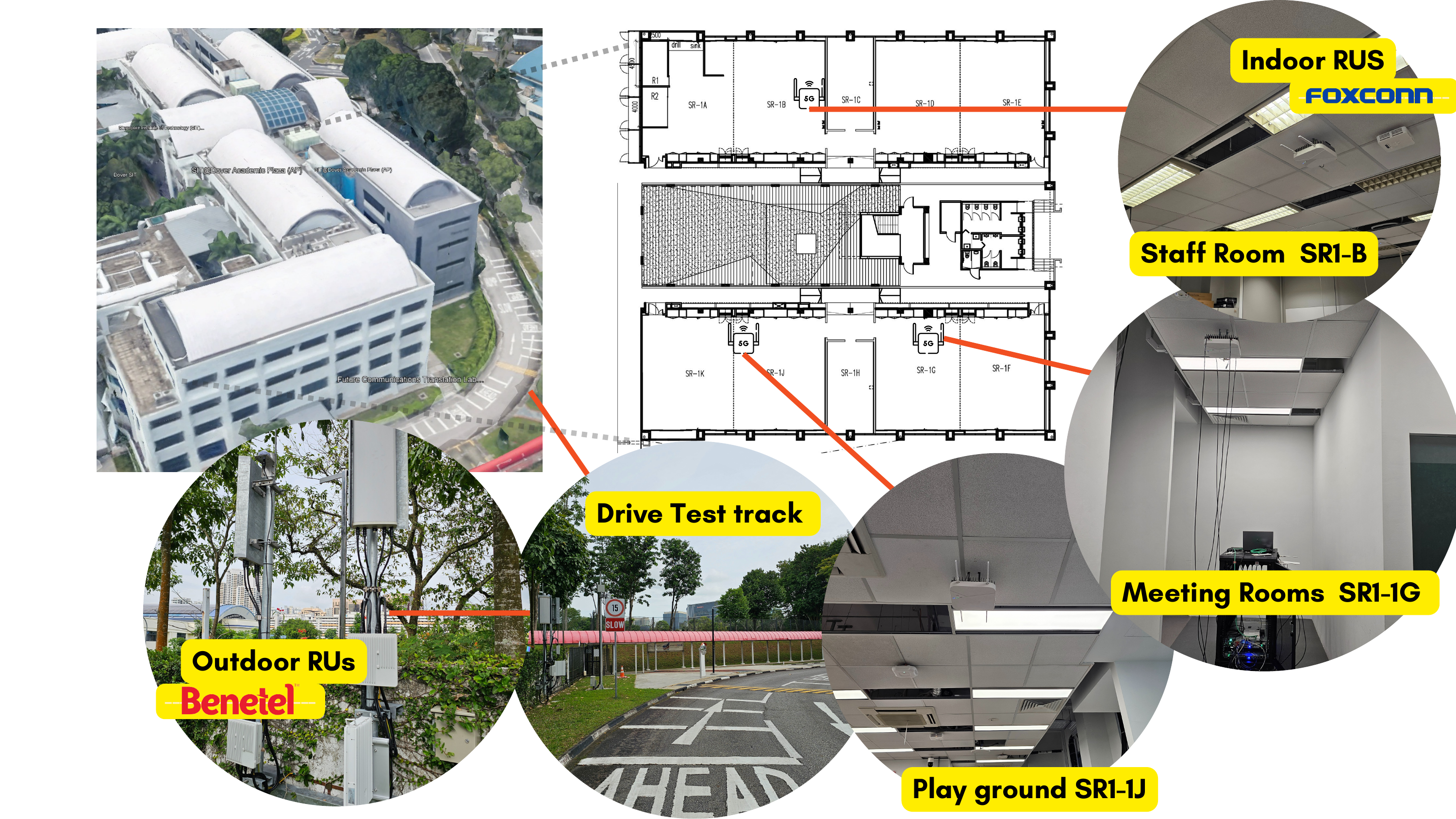}
    \caption{Deployment at SIT: Overview and Infrastructure}
    \label{fig:DeploymentatSIT}
\end{figure}

This section provides an overview of the FCT O-RAN physical deployment currently established at the SIT campus in Dover, located at 10 Dover Drive, Singapore 138683 (Latitude: 1.3006° N, Longitude: 103.7805° E). As illustrated in the Figure~\ref{fig:DeploymentatSIT}, the network setup incorporates both indoor and outdoor radio units to ensure comprehensive and reliable coverage across the designated areas. Three Foxconn RPQN-7800 indoor radio units are deployed to provide seamless connectivity in key indoor locations: the Playground (SR-1J), Meeting Room (SR-1G), and Staff Room (SR-1B).

Two outdoor radio units are positioned to ensure comprehensive coverage along the drive test track, with one oriented towards the Faculty Hall and the other towards the Faculty Block. The configuration parameters for the FCT O-RAN are summarized in Table.\ref{tab:config}.

\begin{table}[h!]
\centering
\begin{tabular}{@{}ll@{}}
\toprule
\textbf{Parameter}                & \textbf{Details}                                               \\ \midrule
\textbf{Frequency Band}           & FR1, n78                                                      \\
\textbf{Bandwidth}                & \begin{tabular}[c]{@{}l@{}}Indoor: 50 MHz \\ Outdoor: 40 MHz\end{tabular} \\
\textbf{NR-ARFCN (SSB Frequency)} & \begin{tabular}[c]{@{}l@{}}Indoor: 627264 \\ Outdoor: 627552\end{tabular} \\
\textbf{RUs}                      & \begin{tabular}[c]{@{}l@{}}Indoor: Foxconn RPQN-7800 \\ Outdoor: Benetel 650\end{tabular} \\
\textbf{Subcarrier Spacing}       & 30 kHz                                                       \\
\textbf{Frequency Range}          & 3400 – 3450 MHz                                              \\
\textbf{PLMN}                     & 00101                                                       \\                                          
\textbf{TDD Slot Configuration}   & DDDSU                                                       \\
\textbf{Central Frequency}        & 3424.980 MHz                                                \\
 \bottomrule
\end{tabular}
\caption{Configuration Parameters for Indoor and Outdoor Details}
\label{tab:config}
\end{table}


\section{Multi-vendor FCT O-RAN Cloud-Native Architecture}
\label{sec:multi_vendor_fct_oran_architecture}

\begin{figure}[h!]
    \centering
    \includegraphics[scale=0.18]{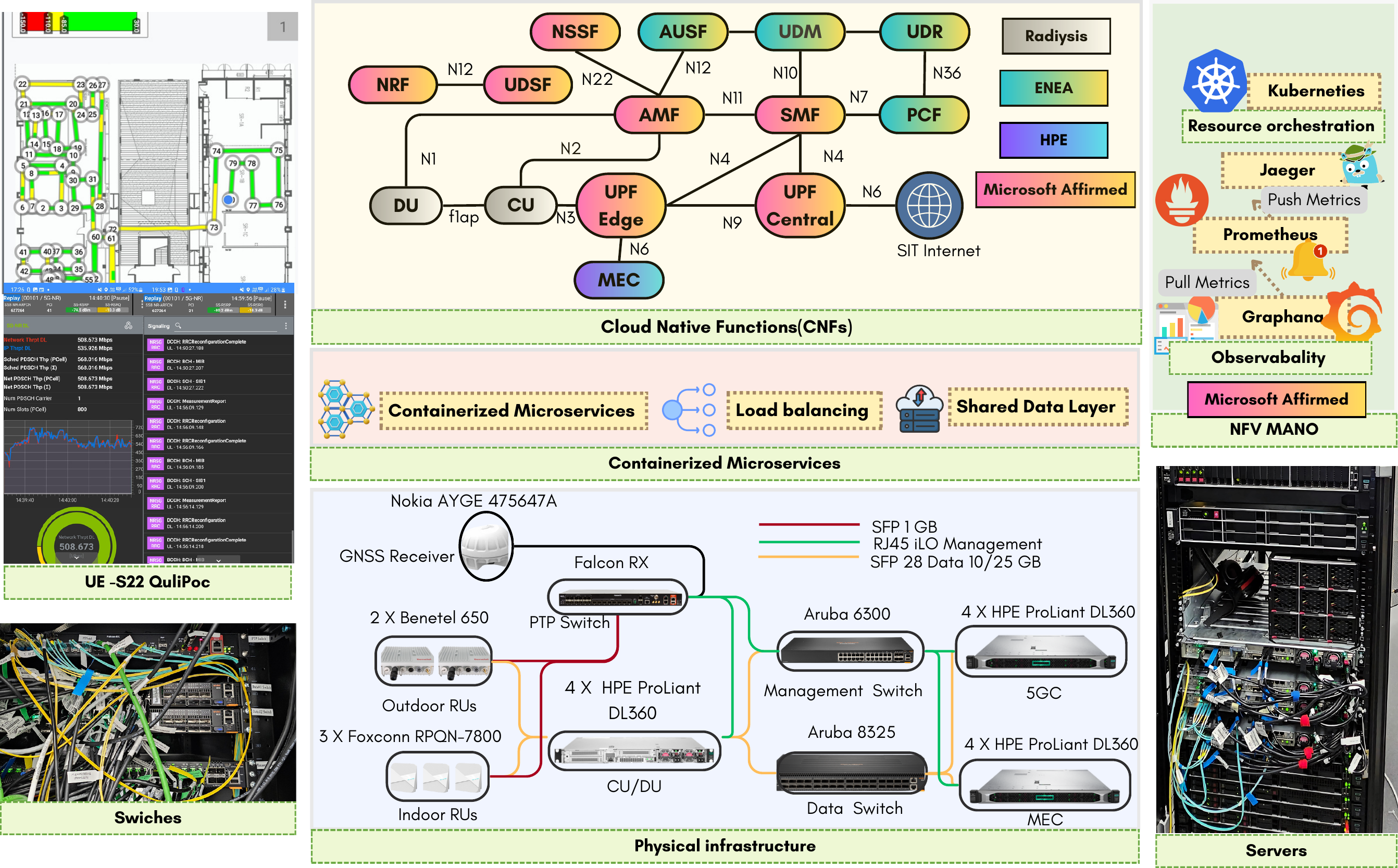}
    \caption{Multi-vendor FCT O-RAN Cloud-Native Architecture}
    \label{fig:multi}
\end{figure}

This section describes the Multi-vendor FCT O-RAN Cloud-Native Architecture, which integrates diverse physical infrastructure components, an ETSI MANO based orchestration framework, and cloud-native 5G network functions. The architecture embodies a modular and scalable design, enabling seamless interoperability across multi-vendor environments. Figure~\ref{fig:multi} illustrates the key elements of this architecture. For performance validation during walk tests, the Samsung S22 device, equipped with QualiPoc$^\text{TM}$, was utilized to conduct comprehensive network evaluations.

\subsection{Physical Infrastructure}

We utilize the Nokia AYE 475647A GNSS module with the Fibrolan Falcon-RX, a 5G xHaul Timing-Aware O-RAN Switch, for synchronization and timing capabilities through SyncE and PTP. The Falcon-RX series features 20 SFP+/SFP28 ports, configurable for up to 200 Gbps full-duplex. We deploy three Foxconn RPQN-7800 indoor 5G NR (New Radio) O-RUs (O-RAN Radio Units) with a 7.2 split architecture. These units support Power over Ethernet (PoE++) via a 10 Gbps Ethernet RJ-45 port. For cells configurations, we use PCI (Physical Cell Identity) 21 for the SR-1B, PCI 31 for the SR-1J, and PCI 41 for the SR-1G. These units feature a compact design that allows flexible installation on ceilings or walls, making them ideal for a variety of indoor environments. For outdoor deployments, we use two Benetel RAN650 O-RUs, which support the 7.2 functional split. Featuring an IP65-rated enclosure, the RAN650 is designed to withstand challenging outdoor conditions and operates with PCI 51 and PCI 61. Additionally, we deploy the Alpha Wireless AW3161 antenna.

We deploy the CU and DU on four HPE ProLiant DL110 Gen10 Plus Telco Servers. These servers deliver exceptional performance through advanced I/O capabilities with PCIe Gen4 speeds, support for hardware acceleration, and powerful computing driven by 3rd Generation Intel\textsuperscript{\textregistered} Xeon\textsuperscript{\textregistered} Processors. Operating at a frequency of 2.3 GHz, these processors feature 20 cores, 30 MB of L3 cache, and support up to 6 TB of DDR4 memory per socket at 2933 MT/s.

We deploy the 5G Core on four HPE ProLiant DL360 Gen10 Servers. Each server is equipped with the 2nd Generation Intel\textsuperscript{\textregistered} Xeon\textsuperscript{\textregistered} Scalable Processor Family, specifically the Intel\textsuperscript{\textregistered} Xeon\textsuperscript{\textregistered} Platinum 8280 Processor. This processor operates at a frequency of 2.7 GHz, features 28 cores, and includes 38.50 MB of L3 cache. It supports three UPI links at 10.4 GT/s and up to 1 TB of DDR4 memory per socket at 2933 MT/s. Additionally, MEC (Multi-access Edge Computing) is deployed on four HPE ProLiant DL360 Gen10 Servers with the same specifications as the 5G Core. These servers include the MEC Orchestrator Server (MEO), the MEC Platform Server (MEP), and two MEC Host Servers (MEH1 and MEH2).

For network management, we use the Aruba 6300 Switch, which offers up to 1,760 Gbps switching capacity, PoE+ and Class 6 PoE support, advanced stacking with 400 Gbps bandwidth, and robust security features such as MACsec encryption and dynamic segmentation, all managed through Aruba Central for efficient operations. For the data switch, we utilize the Aruba 8325, a high-performance switch that provides 48 ports of 25GbE, 8 ports of 100GbE.

\subsection{NFV Management and Orchestation (NFV MANO)}

We leveraged Microsoft Affirmed UnityCloud, a state-of-the-art web-scale, cloud-native NFV MANO platform explicitly designed to align with ETSI standards and support microservices-based Cloud-Native Functions (CNFs). UnityCloud's dynamic orchestration capabilities, powered by Kubernetes, enable the real-time creation, scaling, and removal of microservices. The platform integrates open-source tools, including Jaeger for network monitoring and tracing, and Prometheus along with Grafana for real-time alerting and visualization. Its DevOps-driven environment embraces a CI/CD model, ensuring seamless service automation, efficient operations, and agile development, crucial for the evolving demands of 5G networks.

\subsection{Clould Native Functions} 

\subsubsection{5G NR gNB}

We leverage the O-RAN Radisys Trillium 5G NR gNB solution, which supports both Standalone (SA) and Non-Standalone (NSA) deployment modes. In our current deployment, we utilize the 5G Standalone architecture with the Option 7.2 functional split, integrating the gNB Control Plane (gNB CP) and User Plane (gNB UP) on a single board. This streamlined design optimizes key protocols such as Packet Data Convergence Protocol (PDCP), Service Data Adaptation Protocol (SDAP), and evolved GPRS Tunneling Protocol (eGTP-U) through a scalable worker thread model, improving system efficiency.

\subsubsection{5G Core Network}

We utilize ENEA Integrated 5G Data Management and Policy Control, which includes the Cloud Data Manager, Unified Data Manager (UDM), Unified Data Repository (UDR), and Policy Control Function (PCF), to enable seamless data management and efficient policy control within our 5G core network. This solution supports both standalone and non-standalone 5G deployments. At its foundation, ENEA Stratum serves as a resilient, distributed 5G Network Data Layer (NDL), offering hybrid storage capabilities with both disk-based and in-memory configurations. This flexibility supports structured and unstructured data, catering to a wide range of application needs. The ENEA Unified Data Manager (UDM) complements Stratum by performing essential 3GPP functions, including subscriber data and UE context management, authentication, and event exposure. Policy control is managed by the ENEA PCF, which ensures efficient resource allocation, traffic management, and dynamic QoS adjustments.

We employ Microsoft Affirmed Network Functions, including the Access and Mobility Management Function (AMF), Session Management Function (SMF), User Plane Functions (UPF), Authentication Server Function (AUSF), Network Slice Selection Function (NSSF), Network Repository Function (NRF), and Unified Data Storage Function (UDSF). This architecture supports a robust, scalable, and efficient network that supports diverse applications and use cases.

To optimize latency, scalability, and traffic management, we deploy both Edge and Central UPFs. The Edge UPF, strategically located closer to end users, handles latency-sensitive applications such as immersive multimedia experiences. Its proximity reduces backhaul traffic via local breakout capabilities and accelerates data processing. Seamless integration with Multi-access Edge Computing (MEC)—including the MEC Orchestrator Server (MEO), the MEC Platform Server (MEP), and two MEC Host Servers (MEH1 and MEH2)—enhances localized computing and storage, empowering edge-specific services. The Central UPF, located in the core network, enforces centralized policies and maintains connectivity with our SIT internet data network.

The Access and Mobility Management Function (AMF) manages access and mobility for mobile subscribers. It interfaces with the Radisys Trillium 5G NR gNB through the \textit{N2} interface and handles critical operations such as registration, connection, and mobility management, along with authentication and authorization. The Session Management Function (SMF) is responsible for managing PDU sessions, enabling UEs to access multiple Data Network Names (DNNs). It supports session creation, modification, deletion, network slice selection, QoS maintenance, charging, and policy control while ensuring seamless mobility across the network.

The Network Repository Function (NRF) facilitates NF registration, discovery, and state management through REST APIs, ensuring efficient communication and streamlined operations across the network. The Network Slice Selection Function (NSSF) enhances our testbed by assisting the AMF in allocating slice-specific resources. Supporting inter-Pod and inter-NF communication via REST APIs, the NSSF ensures efficient slice management and isolation. Finally, the Unified Data Storage Function (UDSF) acts as the centralized data repository, storing structured data for efficient access by network functions. This ensures that stateful information is maintained and retrieved seamlessly.

\section{Experimental Results}

\begin{figure}[h!]
    \centering
    \includegraphics[scale=0.15]{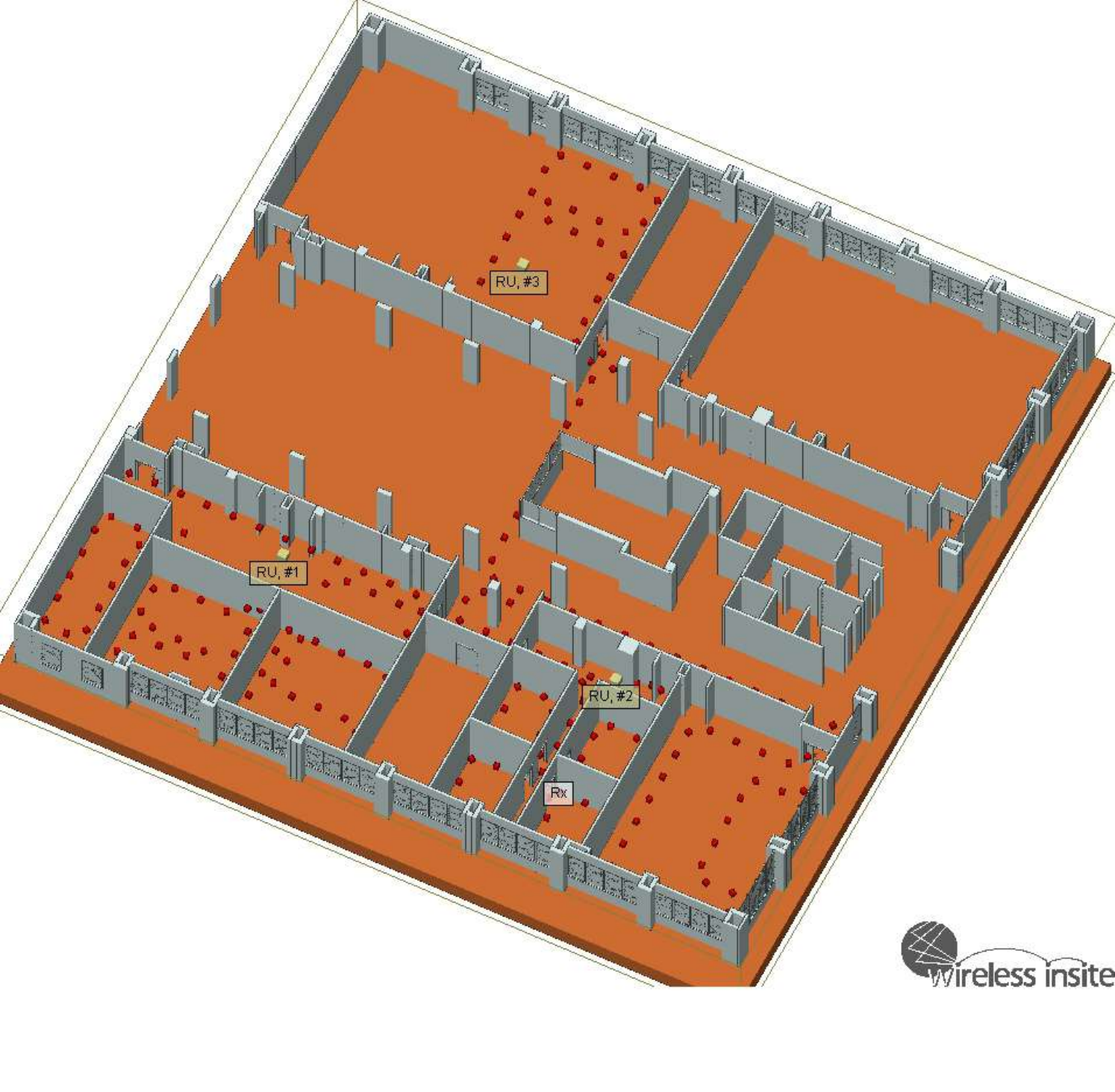}
    \caption{FCTLab Digital Twin Simulation in Wireless InSite}
    \label{3dmodel}
\end{figure}

\begin{figure*}[htb!]
\centering
\minipage{0.32\linewidth}
\centering
\includegraphics[width=0.8\linewidth]{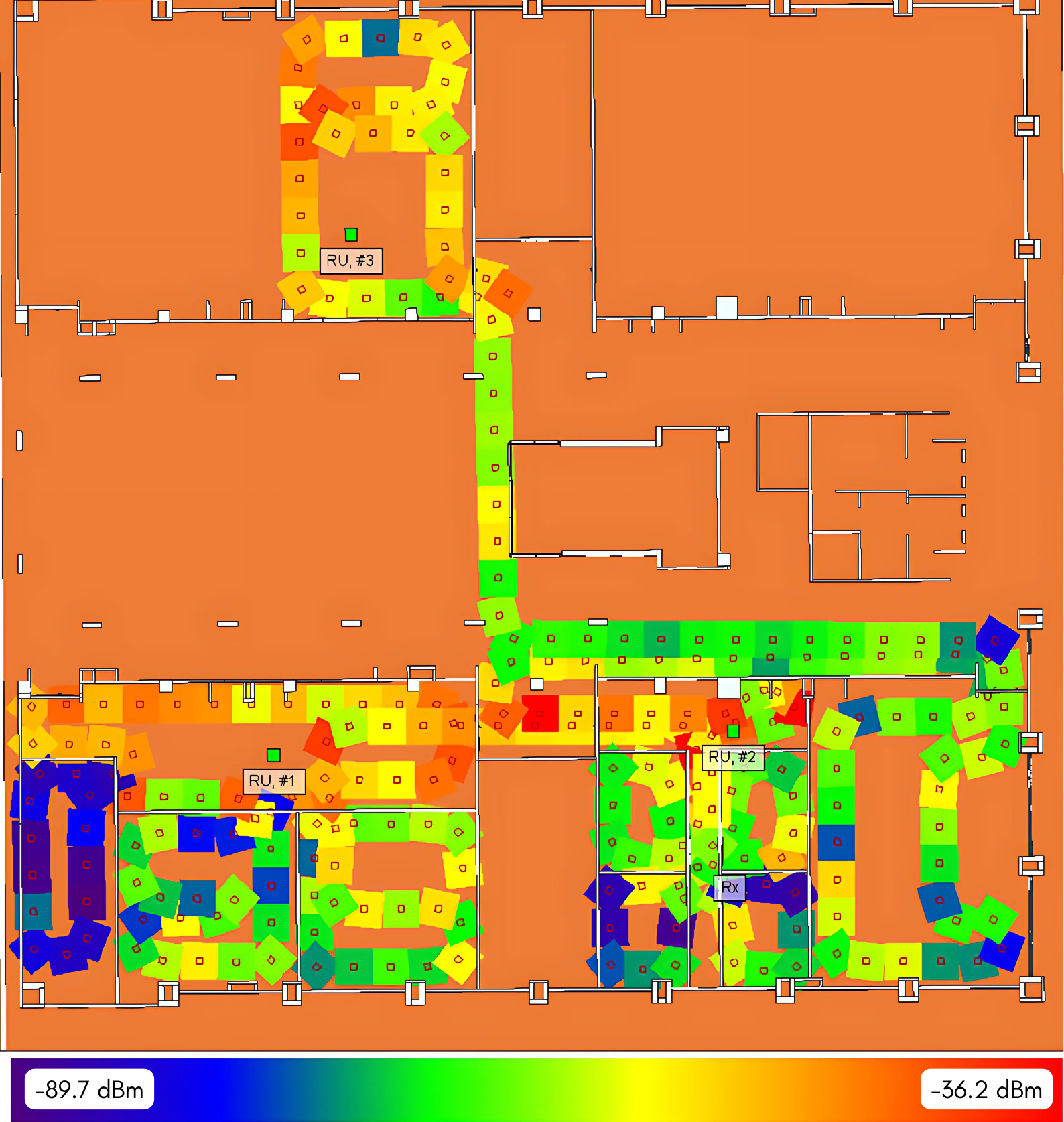}
\caption{RSSI Heatmap from Wireless InSite}
\label{rssi_wi}
\endminipage\hfill
\minipage{0.32\linewidth}
\centering
\includegraphics[width=0.8\linewidth]{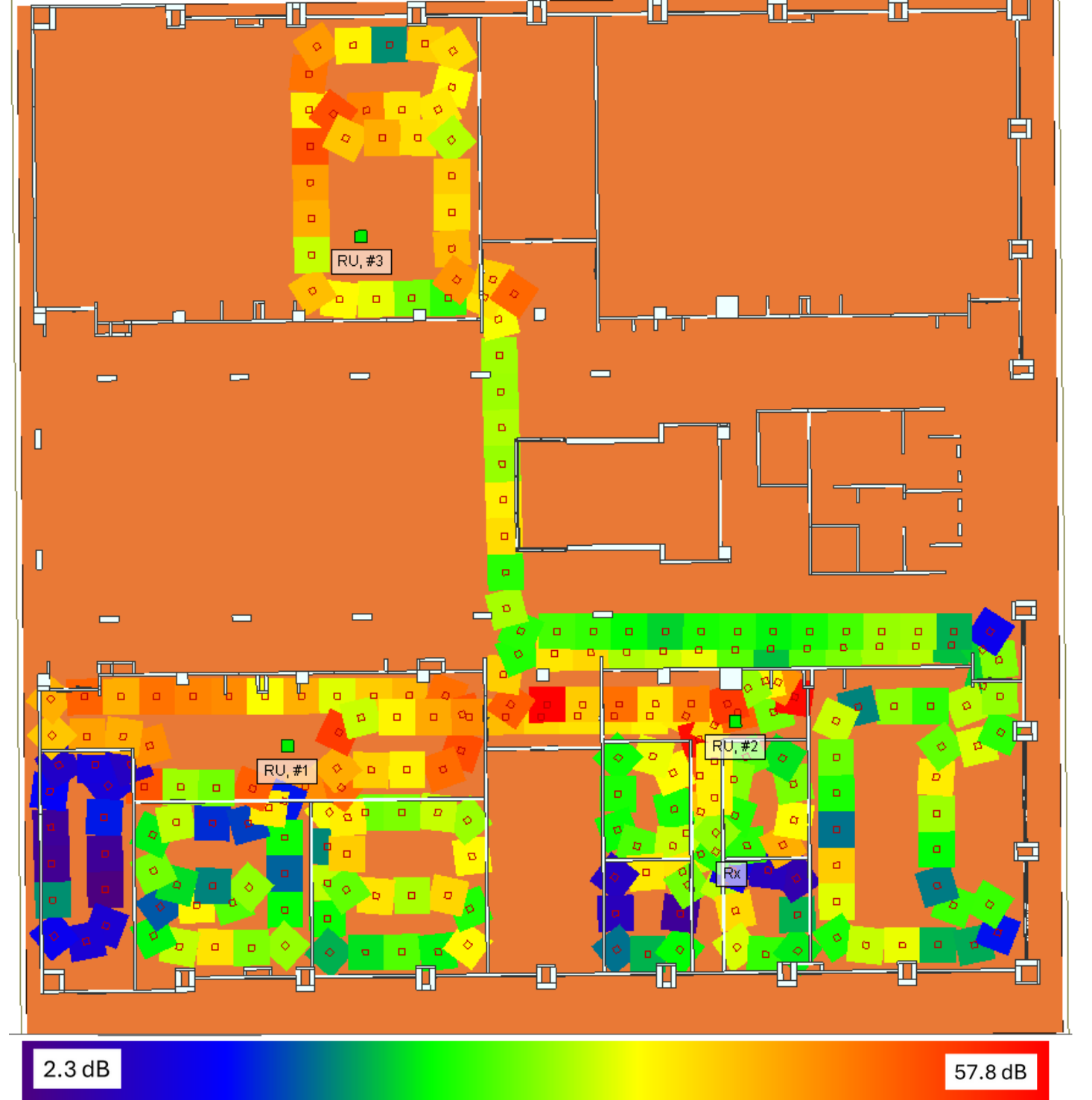}
\caption{SINR Heatmap from Wireless InSite}
\label{sinr_wi}
\endminipage\hfill
\minipage{0.32\linewidth}
\centering
\includegraphics[width=0.8\linewidth]{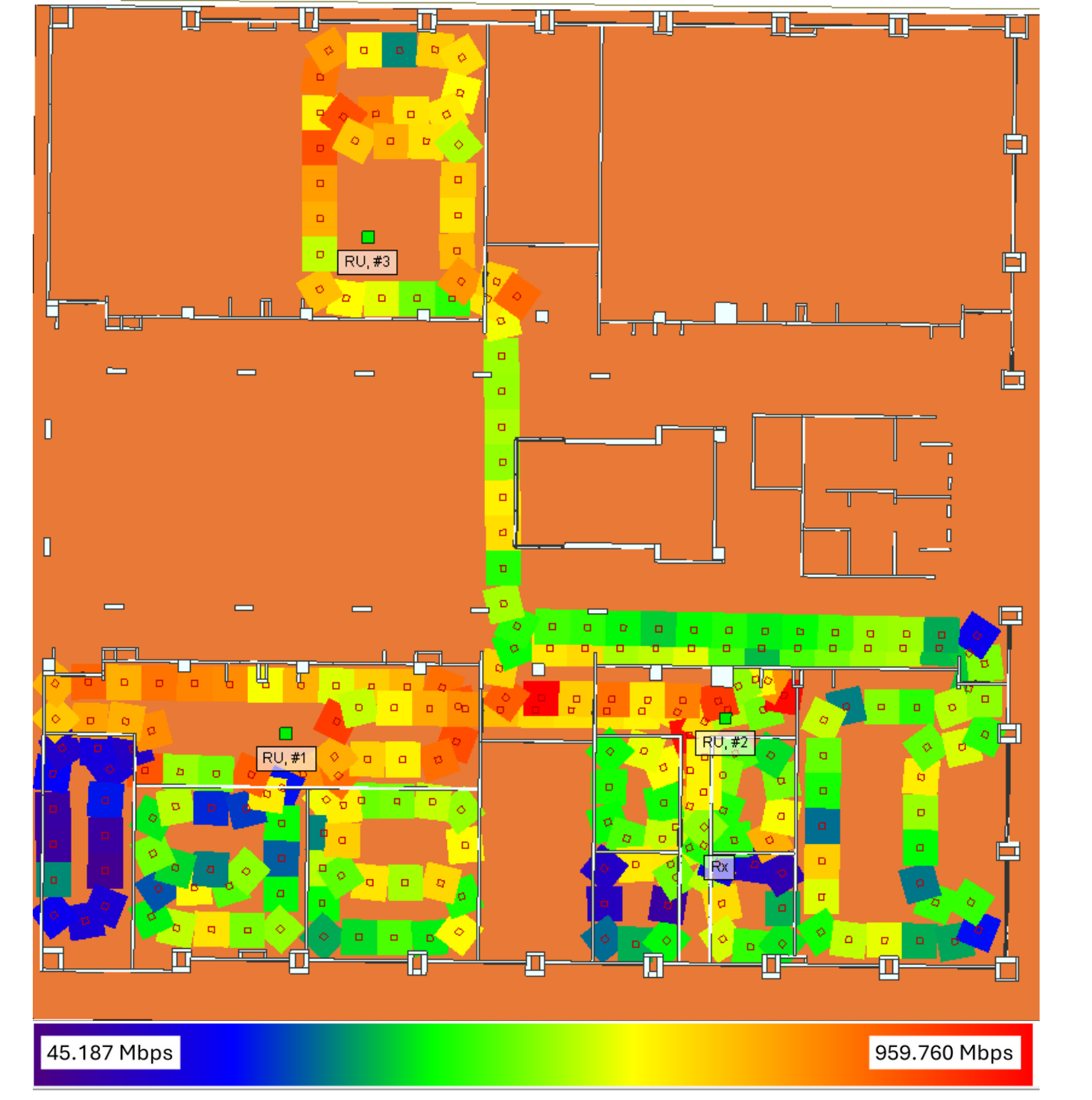}
\caption{Throughput Heatmap from Wireless InSite}
\label{capacity_wi2}
\endminipage\hfill

\vspace{0.3cm}

\minipage{0.32\linewidth}
\centering
\includegraphics[width=0.8\linewidth]{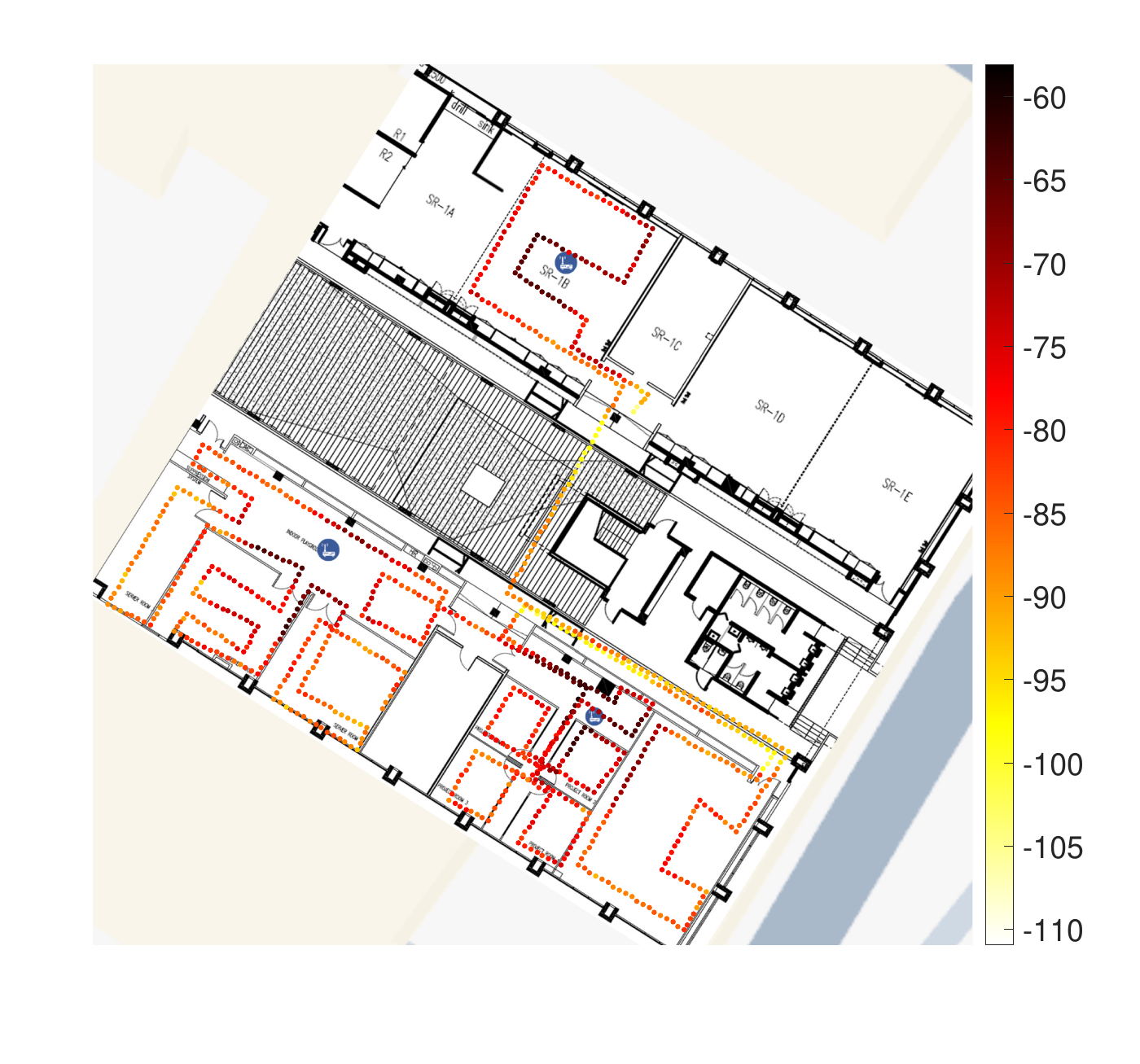}
\caption{RSRP Indoor from QualiPoc$^\text{TM}$}
\label{qualipoc_rsrp}
\endminipage\hfill
\minipage{0.32\linewidth}
\centering
\includegraphics[width=0.8\linewidth]{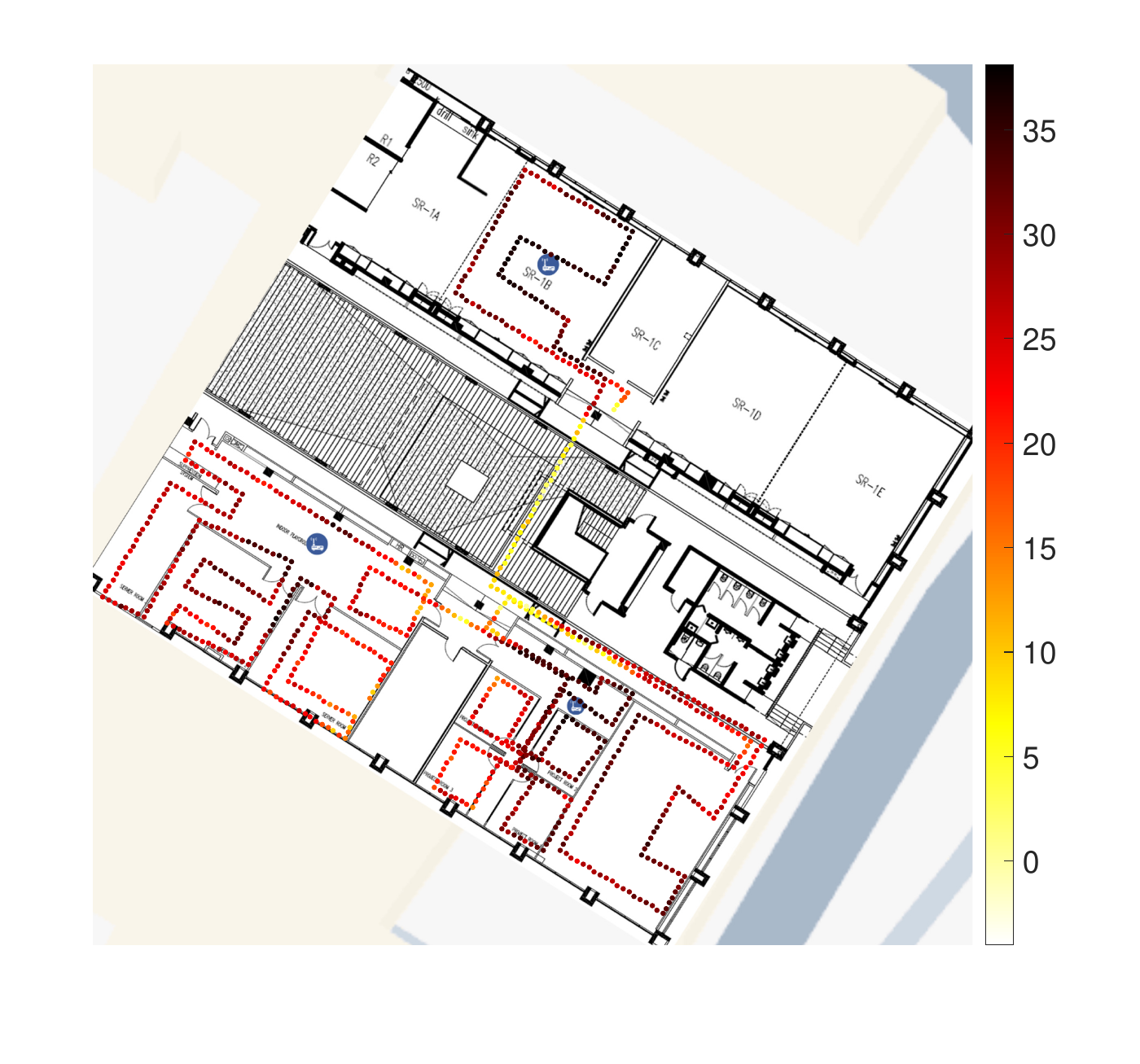}
\caption{SINR Indoor from QualiPoc$^\text{TM}$}
\label{qualipoc_sinr}
\endminipage\hfill
\minipage{0.32\linewidth}
\centering
\includegraphics[width=0.8\linewidth]{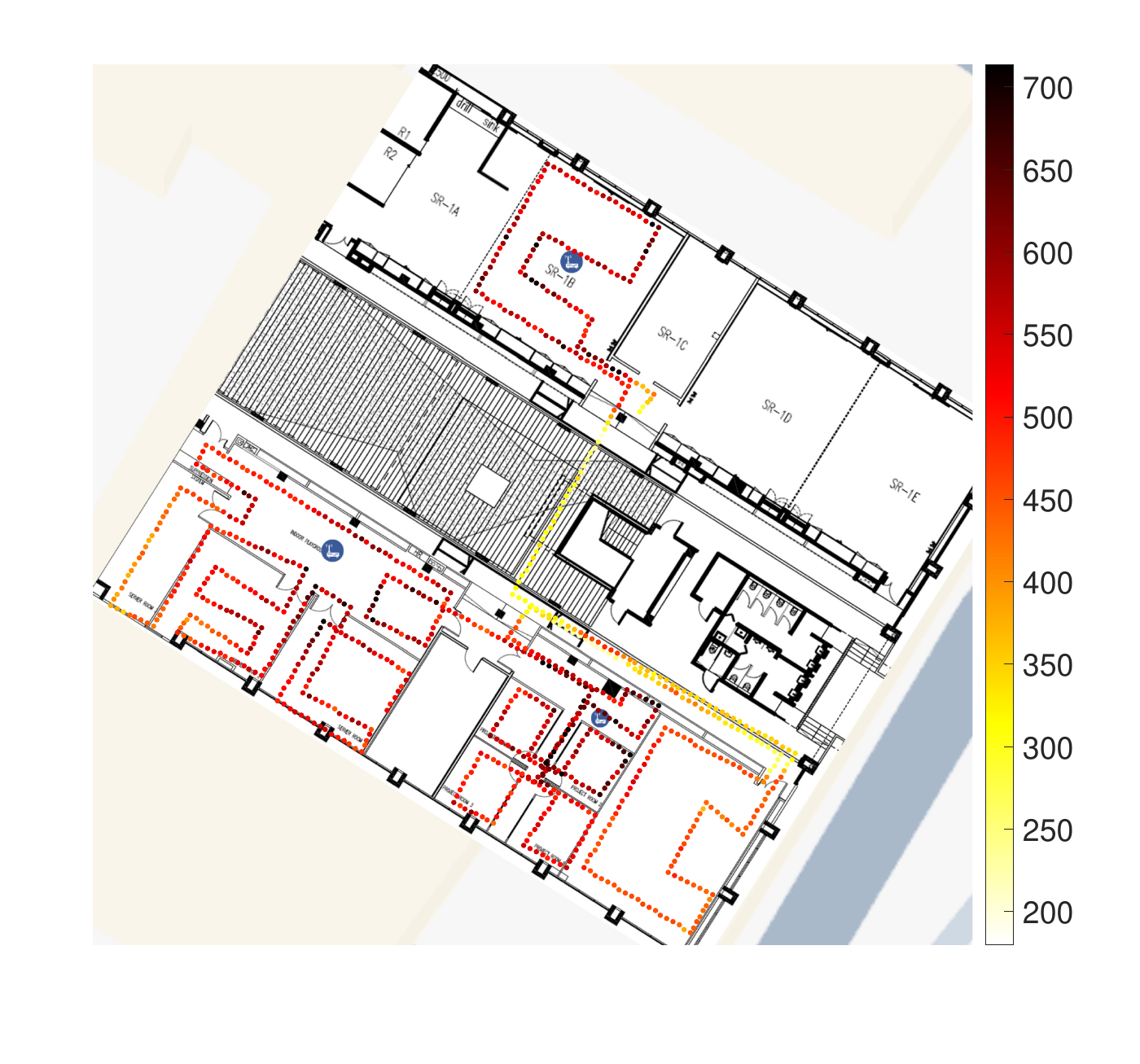}
\caption{Throughput Indoor from QualiPoc$^\text{TM}$}
\label{qualipoc_tput2}
\endminipage
\end{figure*}

\label{sec:experimental_results}
This section discusses the optimization of O-RU placement using a Wireless InSite Digital Twin and the validation of the FCT O-RAN testbed through a walk test. The digital twin is employed to enhance network coverage and performance by simulating and optimizing O-RU placement. Validation is performed using Samsung S22 UEs with the QualiPoc$^\text{TM}$ app, measuring key performance indicators such as throughput, SINR, and RSRP. The real-world measurements are compared with the digital twin's simulation outputs.

The indoor walk test begins at the Playground SR-1J, proceeds through the Meeting Room SR-1G, and concludes in the Staff Room SR-1B. This route is consistently followed for both the Wireless InSite simulation and the Qualipoc$^\text{TM}$ measurements. The real-world data collected during the walk test is subsequently compared to the simulation results generated by the digital twin.

\subsection{Optimization of RF Placement Using Wireless InSite Digital Twin}

A comprehensive digital twin 3D model of the FCT Lab was created in SolidWorks using the building's floor plans. The 3D model was then imported into Wireless InSite Figure \ref{3dmodel} as an STL file to accurately simulate the propagation environment. In \cite{10388951}, the use of 3D ray-tracing simulations incorporating a multi-wall model was presented for RU placement optimization. In this work, we utilize the 3D model to create a digital twin of the network deployment.

The parameters used for the ray-tracing simulation are tabulated in Table \ref{table:simulation_features}. These parameters are configured to reflect the physical and operational characteristics of the deployment, including material properties, antenna patterns and configurations. Additionally, the X3D propagation model in Wireless InSite is utilized.

\begin{table}[h!]
\centering
\caption{Wireless InSite Simulation}
\label{table:simulation_features}
\begin{tabular}{@{}ll@{}}
\toprule
\textbf{Parameter}                & \textbf{Details}   \\ \midrule
\textbf{Materials}                & Concrete walls, thickness 30 cm \\  
\textbf{Frequency Band}           & FR1, n78           \\
\textbf{Carrier Frequency}        & 3.424 GHz           \\
\textbf{Bandwidth}                & 50 MHz               \\
\textbf{Components Used}  & 3 TX, 250 RX         \\
\textbf{Types of Antennas Used}   & \begin{tabular}[c]{@{}l@{}}TX: 4x4 Rectangular Patch \\ RX: Halfwave Dipole\end{tabular}     \\
\textbf{Transmit Power}           & 20 dBm                \\
\textbf{Transmitter Gain}         & 5 dBi                 \\
\textbf{Propagation Model}        & X3D                   \\
\textbf{Noise}                    & -102 dBm              \\
\textbf{Study Area Properities}   & 6 Reflections, 2 Transmissions, 1 Diffractions                    \\
\textbf{Wireless Access Method }  & 5G NR FR1             \\
\bottomrule
\end{tabular}
\end{table}

The receiver locations are indicated as red cubes in the 3D model in Figure \ref{3dmodel} and are placed along the physical walk test route. Following Remcom's recommendation \cite{WirelessInsiteManual}, the simulation accounts for six reflections, one diffraction, and two transmissions to provide realistic results.

The transmitters, labeled as RU1, RU2, and RU3, are placed at the Staff Room, Playground, and Meeting Room, respectively, based on the digital twin’s optimization insights. The tool predicts Reference Signal Strength Indicator (RSSI), SINR, and throughput values for the given simulation setup. The heat maps for RSSI, SINR, and throughput are shown in Figures \ref{rssi_wi}, \ref{sinr_wi}, and \ref{capacity_wi2} respectively.

Figure \ref{rssi_wi} and \ref{capacity_wi2} shows that RSSI along the route of walk test varies  from -89.7 dBm to -36.2 dBm. Throughput varies from 45 Mbps to 959 Mbps. Along the corridor, RSSI varies from -68 dBm to -55 dBm, with throughput between 387 Mbps and 583 Mbps. Line Of Sight (LOS) locations have RSSI below -55 dBm and throughput above 630 Mbps. Non-Line Of Sight (NLOS) locations have RSSI below -80 dBm and throughput below 220 Mbps. The simulations predicts a signal loss of at least 10 dB for each wall the signal passes through.

\begin{figure*}[htb!]
\centering

\minipage{0.32\textwidth}
\centering
\includegraphics[width=1\textwidth]{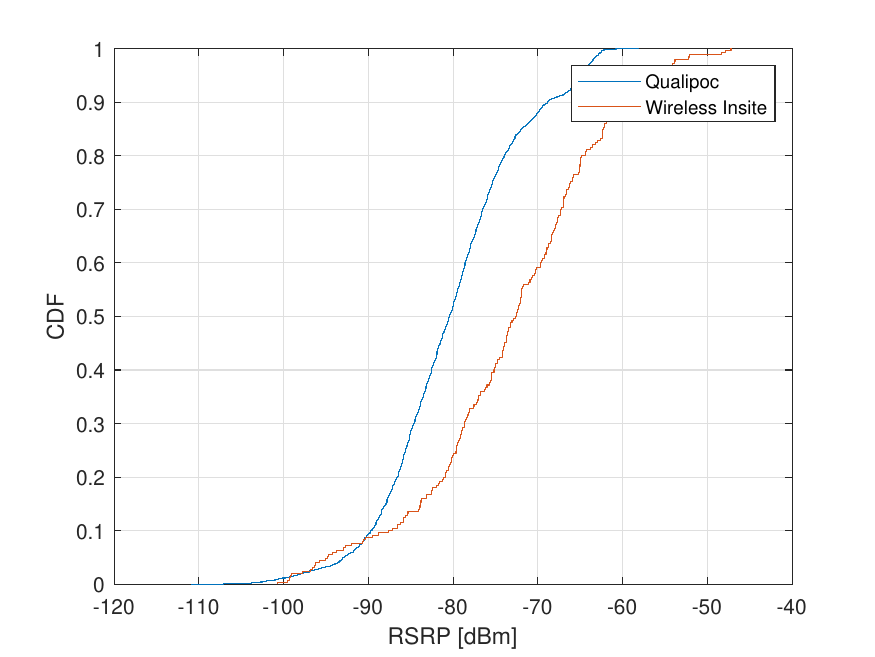}
\caption{CDF Comparison of Indoor RSRP}
\label{rx_cdf_compare}
\endminipage\hfill
\minipage{0.32\textwidth}
\centering
\includegraphics[width=1\textwidth]{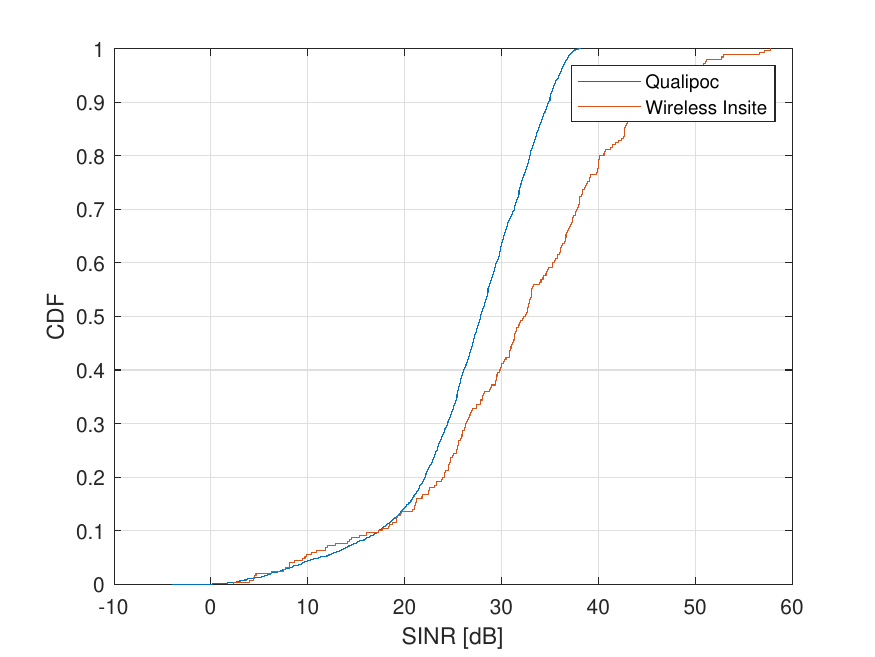}
\caption{CDF Comparison of Indoor SINR}
\label{sinr_cdf_compare}
\endminipage\hfill
\minipage{0.32\textwidth}
\centering
\includegraphics[width=1\textwidth]{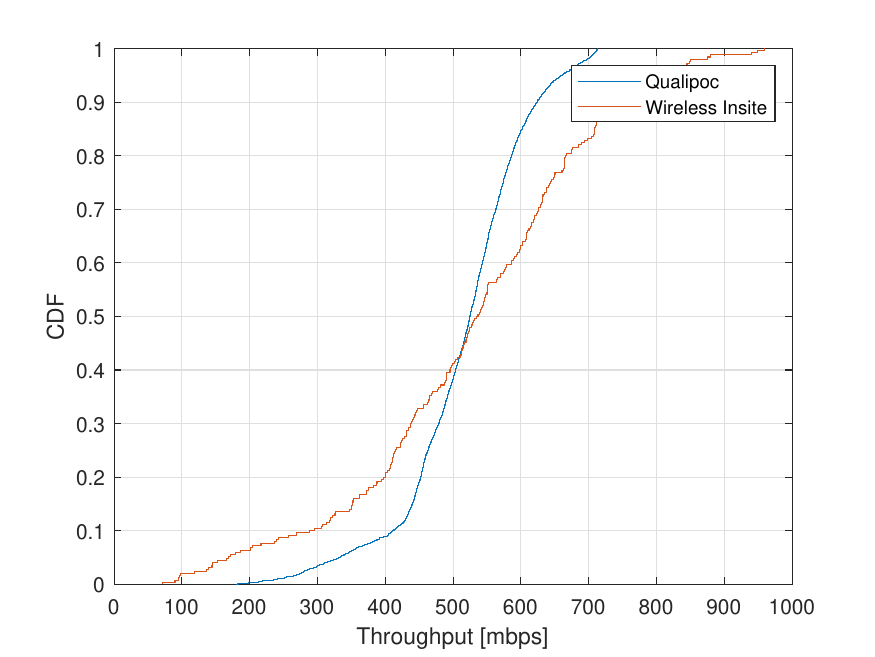}
\caption{CDF Comparison of Indoor Throughput}
\label{tput_cdf_compare}
\endminipage

\end{figure*}

\subsection{Measurement Campaign using QualiPoc$^\text{TM}$}

A walk test was conducted with Samsung S22 along the route to validate the performance of the FCT O-RAN testbed. Being an indoor test,  the floor plan of the FCT Lab is uploaded and overlaid on a Google Maps image within the QualiPoc$^\text{TM}$ app. Since GPS is not available indoors, we manually place static points on the map to log and save the measurement data accurately. After completing the walk test with QualiPoc$^\text{TM}$, the measurement data is saved, and heatmaps for key performance indicators, including SS-RSRP, SINR, and Throughput, are generated. These heatmaps are presented in Figures \ref{qualipoc_rsrp}, \ref{qualipoc_sinr}, and \ref{qualipoc_tput2}. For reference, the locations of the three RUs are marked with a blue antenna symbol. Figures \ref{qualipoc_rsrp} and \ref{qualipoc_tput2} show that in the indoor environment, the SS-RSRP along the entire route varies from -110 dBm to -58 dBm, with downlink throughput ranging from 179 Mbps to 713 Mbps. Along the corridor, SS-RSRP ranges from -101 dBm to -82 dBm, depending on proximity to rooms with O-RUs, and downlink throughput varies between 266 Mbps and 458 Mbps. LOS locations have SS-RSRP values above -80 dBm, with downlink throughput exceeding 500 Mbps. In NLOS areas, SS-RSRP drops below -85 dBm, and throughput falls below 400 Mbps. With the Samsung S22 operating on a single-layer uplink antenna, we achieved an uplink throughput of 35 Mbps, whereas the Netgear CPE, equipped with a dual-uplink antenna, achieved 66 Mbps, demonstrating the performance benefits of dual-layer transmission.

Due to space constraints, the results for the outdoor measurement results are not included. However, our analysis indicates that a Line-of-Sight (LOS) connection is achieved at most locations outdoors. The observed SS-RSRP values range from -85 dBm to -43 dBm, with downlink throughput varying between 82 Mbps and 371 Mbps. Notably, the maximum throughput in the outdoor environment is lower than that in the indoor environment. This reduction is primarily attributed to the smaller bandwidth and wider beamwidth used by the outdoor O-RUs. Additionally, the outdoor environment has fewer signal reflections, which reduces the multipath effect and further contributes to the lower throughput observed. Outdoors, the Samsung S22 achieved an uplink throughput of 26 Mbps, while the Netgear CPE reached 55 Mbps.

\subsection{Comparision of Wireless InSite and QualiPoc$^\text{TM}$ Results}

By comparing Figures \ref{rssi_wi} and \ref{qualipoc_rsrp}, both simulations and QualiPoc$^\text{TM}$ measurements demonstrate the lowest received signal power along the outside corridor and in NLOS areas, such as the server room in the playground. In contrast, LOS areas, such as rooms with O-RUs, exhibit higher received signal power. Both figures indicate an approximate 10 dB loss when the signal passes through a wall, with the magnitude of the loss varying depending on the angle of incidence.

Figures \ref{sinr_wi} and \ref{qualipoc_sinr} follow the same trend as received power, with SINR decreasing with obstacles or distance from RUs. As shown in Figures \ref{capacity_wi2} and \ref{qualipoc_tput2}, the simulation shows a higher range of throughput than the QualiPoc$^\text{TM}$ measurements, likely because changes in SINR have a more significant impact in the simulation compared to real-life conditions.

Simulation results provide RSSI, whereas QualiPoc$^\text{TM}$ measurements yield SS-RSRP. SS-RSRP represents the linear average of the received power of the Secondary Synchronization Signals (SSS) levels. RSSI measures the overall received signal power over the selected bandwidth, including noise, interference, and the desired signal. To align these metrics, we have converted RSSI to SS-RSRP. The SS-RSRP can be calculated from the RSSI using the following equation, where \(\mathbf{N_{SSS}}\) denotes the number of subcarriers in SSS and \(\mathbf{N_{RB}}\) is the total number of RBs.

\begin{equation}
\text{SS-RSRP} = \text{RSSI} + 10 \cdot \log_{10}\left(\frac{N_{SSS}} {12 \cdot N_{RB}} \right)
\label{eq:ss_rsrp_combined}
\end{equation}

From the SSB block diagram, the SSS consists of 127 subcarriers, and the whole SSB consists of 960 subcarriers. For a bandwidth of 50 MHz with a subcarrier spacing of 30 kHz, \(\mathbf{N_{RB}}\) equals 133, excluding guard bands. Since each RB contains 12 subcarriers, the total number of subcarriers is 1596. The cumulative distribution function (CDF) of RSRP and SINR was plotted to compare the predicted results from Wireless InSite with the QualiPoc$^\text{TM}$ measurements, as shown in Figures \ref{rx_cdf_compare} and \ref{sinr_cdf_compare}.

As shown in Figure \ref{rx_cdf_compare}, the CDF of RSRP from the simulation and QualiPoc$^\text{TM}$ overlap at low RSRP, but the difference increases at higher RSRP values. Higher RSRP or SINR typically corresponds to a LOS scenario, where antenna characteristics significantly affect performance, causing greater discrepancies between simulated and measured results. Figure \ref{tput_cdf_compare} shows a strong correlation between the throughput CDFs from Wireless InSite and QualiPoc$^\text{TM}$ measurements. Therefore, it can be concluded that Wireless InSite accurately represents the real-life O-RAN setup that was measured with QualiPoc$^\text{TM}$.

\section{Conclusion \& Future Work}
\label{sec:conclusion}

This paper presents FCT O-RAN, a multi-vendor private 5G testbed integrating Microsoft Affirmed, ENEA, Radisys, Foxconn, and Benetel, while leveraging Wireless InSite for RF optimization. The testbed achieved impressive performance metrics, including high throughput and low latency, in both indoor and outdoor scenarios. Future work will focus on demonstrating advanced 5G capabilities within the FCT O-RAN platform, including the implementation of network slicing, MEC, and translational 5G use cases across diverse industry scenarios.

\section*{Acknowledgment}
This research is supported by the National Research Foundation, Singapore and Infocomm Media Development Authority under its Future Communications Research \& Development Programme.

\end{document}